\documentclass[sigconf]{acmart}
\usepackage{booktabs} 
\usepackage{hyperref}
\usepackage{relsize}

\usepackage{url}
\usepackage{graphicx}
\usepackage{amssymb}
\usepackage{multirow}
\usepackage{mathtools}
\usepackage{subcaption}
\usepackage[para]{footmisc}
\usepackage{fnpct}
\usepackage{scalerel} 
\usepackage{bbm} 
\usepackage{xcolor,soul} 

\setcopyright{rightsretained}
\acmConference[WSDM]{The 12th ACM International Conference on Web Search and Data Mining}
\acmYear{2019}
\copyrightyear{2018}
\acmISBN{}

\renewcommand\footnotetextcopyrightpermission[1]{} 
\begin{document}
\title{Neural Cross-Domain Collaborative Filtering with Shared Entities}

\author{Vijaikumar M}
\authornote{Corresponding author}
\affiliation{%
  \institution{Indian Institute of Science}
  \city{Bangalore}
  \country{India}}
\email{vijaikumar@iisc.ac.in}
\author{Shirish Shevade}
\affiliation{%
  \institution{Indian Institute of Science}
  \city{Bangalore}
  \country{India}}
\email{shirish@iisc.ac.in}
\author{M N Murty}
\affiliation{%
  \institution{Indian Institute of Science}
  \city{Bangalore}
  \country{India}}
\email{mnm@iisc.ac.in}

\begin{abstract}
Cross-Domain Collaborative Filtering (CDCF) provides a way to alleviate data sparsity and cold-start problems present in recommendation systems by exploiting the knowledge from related domains. Existing CDCF models are either based on matrix factorization or deep neural networks. Either of the techniques in isolation may result in suboptimal performance for the prediction task. Also, most of the existing models face challenges particularly in handling diversity between domains and learning complex non-linear relationships that exist amongst entities (users/items) within and across domains. In this work, we propose an end-to-end neural network model -- NeuCDCF, to address these challenges in cross-domain setting. More importantly, NeuCDCF follows wide and deep framework and it learns the representations combinedly from both matrix factorization and deep neural networks. We perform experiments on four real-world datasets and demonstrate that our model performs better than state-of-the-art CDCF models.
\end{abstract}

\keywords{Cross-domain collaborative filtering, Deep learning, Neural networks, Wide and deep framework, Recommendation system}

\maketitle
\section{Introduction}
\label{sec:intro}
Personalized recommendation systems play an important role in extracting relevant information for user requirements, for example, product recommendation from Amazon\footnote{www.amazon.com}$,$ event recommendation from Meetup\footnote{www.meetup.com}$,$ scholarly references recommendation from CiteULike\footnote{www.citeulike.org}$.$ Collaborative Filtering (CF) is one of the widely used techniques \cite{su2009survey} in recommendation systems that exploits the interactions between users and items for predicting unknown ratings.

Performance of the CF techniques mainly depends on the number of interactions the users and items have in the system. However, in practice, most of the users interact with very few items. For instance, even the popular and well-preprocessed datasets such as Netflix\footnote{https://www.kaggle.com/netflix-inc/netflix-prize-data} and MovieLens-20M\footnote{http://files.grouplens.org/datasets/movielens} have 1.18\% and 0.53\% available ratings, respectively. Furthermore, new users and items are added to the system continuously. Due to this, the new entities (users/items) may have a very few ratings associated with them, and oftentimes, they may not have any ratings. The former issue is called data sparsity problem and the latter is called cold-start problem \cite{su2009survey}; the respective entities are called cold-start entities. When the above-mentioned issues exist, learning parameters in CF techniques becomes a daunting task.

Cross-Domain Collaborative Filtering (CDCF) \cite{cantador2015cross,khan2017cross} is one of the promising techniques in such a scenario that alleviates data sparsity and cold-start problems by leveraging the knowledge extracted from user-item interactions in related domains\footnote{we follow the domain definition as in \cite{lian2017cccfnet,man2017cross,li2009transfer}}$,$ where information is rich, and transfers appropriately to the target domain where a relatively smaller number of ratings are known. In this work, we study the problem of Cross-Domain Recommendation (CDR) with the following assumption: users (or items) are shared across domains and no side information is available. 

Most of the existing models proposed for CDR are based on Matrix Factorization (MF) techniques~\cite{he2018robust,sahebi2015takes,gao2013cross,li2009can,li2009transfer,singh2008relational}. They may not handle \textbf{complex non-linear relationships} that exist among entities within and across domains very well. More recently, few works have been proposed for cross-domain settings to exploit the transfer learning ability of the deep networks \cite{elkahky2015multi,lian2017cccfnet,man2017cross,zhu2018deep}. Although they have achieved performance improvement to some extent, they have the following shortcomings. First, they do not consider the \textbf{diversity} among the domains. That is, they learn shared representations from the source and target domain together instead of learning domain-specific and domain-independent representations separately. Second, they do not consider both the \textbf{wide} (MF based) \textbf{and deep} (deep neural networks based) representations together. We explain why addressing the aforementioned shortcomings are essential in the following:\vspace{0.3cm}\\
\noindent
\textbf{Importance of domain-specific embeddings to handle diversity.}
When users (or items) are shared between different domains one can ignore the diversity among domains and treat all the items (or users) come from the same domain, and apply the single domain CF models. However, it loses important domain-specific characteristics that are explicit to different domains. For example, let us assume that we have two domains: movie and book. Suppose the movie domain is defined by the features: \textit{genre}, \textit{language} and \textit{visual effects}; and the book domain is defined by the features: \textit{genre}, \textit{language} and \textit{number of pages}. Though the features \textit{genre} and \textit{language} are shared by both the domains, \textit{visual effects} and \textit{number of pages} are specific to the movie and book domains, respectively.

\begin{figure}[h]
    \centering
    \includegraphics[width=0.35\textwidth]{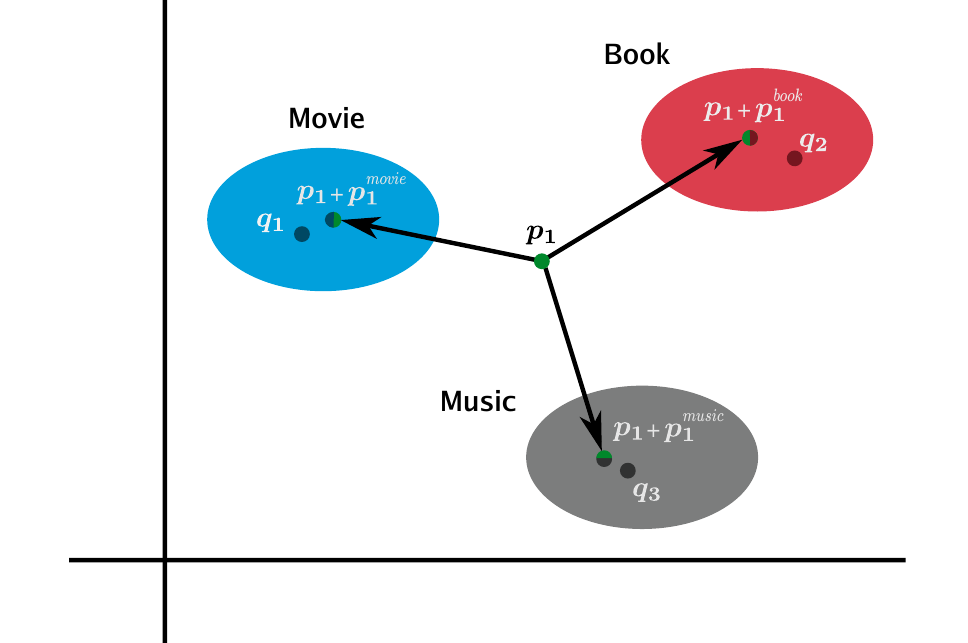} 
    \vspace{-0.45cm}
    \caption{Illustration of domain-specific preferences for cross domain setting to handle diversity.}
    \label{fig:user_rep}
\end{figure}

When users and items are embedded in some latent space, learning only common preferences of users may not be sufficient to handle diversity among domain. 
This is illustrated in Figure \ref{fig:user_rep}. Suppose, user $u1$ rates high on items $j1,j2$ and $j3$ belonging to movie, book and music domains, respectively. Let $p_1, q_1, q_2$ and $q_3$ be embeddings for user $u1$ and items $j1,j2$ and $j3$, respectively. In order to maintain higher similarity values with all the items $j1, j2$ and $j3$, the existing models \cite{he2017neural,koren2009matrix,mnih2008probabilistic,singh2008relational} embed $p_1$ for user $u1$ as shown in Figure \ref{fig:user_rep}. 
This may deteriorate the performance of the prediction task for CDR. This is because we need to account for the domain-specific embeddings $p_1^{movie}$, $p_1^{book}$ and $p_1^{music}$ to get the appropriate preferences for movie, book and music domains, respectively.
\vspace{0.3cm}\\
\noindent
\textbf{Importance of learning non-linear relationships across domains.}
Learning non-linear relationships present across domains is important in transferring knowledge across domains. For example\footnote{This example is inspired from \cite{liu2015non}.}, let us assume that a movie production company releases a music album in advance to promote a movie. Let one feature for `music' be \textit{popularity}, and the two features for `movie' be \textit{visibility} (reachability to the people) and \textit{review-sentiment}. As the \textit{popularity} of the music album increases, the \textit{visibility} of the movie also increases and they are positively correlated. However, the movie may receive positive reviews due to the music \textit{popularity} or it may receive negative reviews if it does not justify the affiliated music album's \textit{popularity}. So, the movie's \textit{review-sentiment} and music album's \textit{popularity} may have a non-linear relationship, and it has been argued \cite{he2017neural,rendle2010factorization} that such relationships may not very well be handled by matrix factorization models. Thus, it is essential to learn non-linear relationships for the entities in cross-domain systems using deep networks.\vspace{0.3cm}\\
 \noindent
\textbf{Necessity of fusing wide and deep networks.}
 The existing models for CDR are either based on MF techniques or neural network based techniques. However, following either approach leads to suboptimal performance. That is, deep networks are crucial to capture the complex non-linear relationships that exist between entities in the system. At the same time, we cannot ignore the complementary representations provided by wide networks such as matrix factorization~\cite{cheng2016wide}.\vspace{0.3cm}\\
\noindent
\textbf{Contributions.} In this work, we propose a novel end-to-end neural network model for CDR, \underline{Neu}ral \underline{C}ross-\underline{D}omain \underline{C}ollaborative \underline{F}iltering (NeuCDCF), to address the above-mentioned challenges. Our novelty lies particularly in the following places:
\begin{itemize}
    \item We extend MF to CDR setting to learn the wide representations. We call this MF extension as Generalized Collective Matrix Factorization (GCMF). We propose an extra component in GCMF that learns the domain-specific embeddings. This addresses the diversity issues to a great extent.
    \item Inspired from the success of encoder-decoder architecture in machine translation~\cite{cho2014learning,bahdanau2014neural,cho2014properties}, we propose Stacked Encoder-Decoder (SED) to obtain the deep representations for the shared entities. These deep representations learn the complex non-linear relationships that exist among entities. This is obtained by constructing target domain ratings from source domain.
    \item More importantly, we fuse GCMF and SED to get the best of both wide and deep representations. All the parameters from both the networks are learned in end-to-end fashion. To our best knowledge, NeuCDCF is the first model that is based on wide and deep framework to CDR setting.
\end{itemize}
In addition, GCMF generalizes many well-known MF models such as PMF \cite{mnih2008probabilistic}, CMF\cite{singh2008relational} and GMF\cite{he2017neural}.
Further, source-to-target rating construction part of SED acts as a regularizer for CDR setting.
We conduct extensive experiments on four real-world datasets -- two from Amazon and two from Douban, in various sparse and cold-start settings, and demonstrate the effectiveness of our model compared to the state-of-the-art models. Our implementation will be made available online. 

\section{Problem Formulation}
\label{sec:prob}
We denote the two domains, source and target, by the superscripts $S$ and $T$, respectively. Ratings from the source and target domains are denoted by the matrices $R^S = [r_{uj}^S]_{m\times n_S}$ and $R^T = [r_{uj}^T]_{m\times n_T}$ where $r_{uj}^S,r_{uj}^T \in [\gamma_{min},\gamma_{max}] \cup \{0\}$, $u$ and $j$ denote user and item indices, $0$ represents unavailability of rating for the pair $(u,j)$, and $\gamma_{min}$ and $\gamma_{max}$ denote minimum and maximum rating values in rating matrices $R^S$ and $R^T$, respectively. Further, $m,n_S$ and $n_T$ represent the number of users, the number of items in source domain and the number of items in target domain, respectively. Let, $n = n_S + n_T$. We indicate the available ratings in source domain by $\Omega^S = \{(u,j)| r_{uj}^S \neq 0\}$ and target domain by $\Omega^T = \{(u,j)| r_{uj}^T \neq 0\}$. We drop the superscripts $S$ and $T$ from $r_{uj}^S$ and $r_{uj}^T$ wherever the context is clear through $\Omega^S$ and $\Omega^T$.

Further, $p \odot q$ and $p'q$ denote element-wise multiplication and dot product of two vectors $p$ and $q$ of the same size, $a(\cdot)$ denotes activation function, and $\| x \|$ denotes $l_2$ norm of a vector $x$. Let $R_{u,:}$ ($R_{:,j}$) be $u\textsuperscript{th}$ row ($j\textsuperscript{th}$ column) of the matrix $R$ corresponding to the user $u$ (item $j$).

A set of users, shared across domains, is denoted by $\mathcal{U}$ and the sets of items for source and target domains are denoted by $\mathcal{V}^S$ and $\mathcal{V}^T$, respectively. Let $\mathcal{V} = \mathcal{V}^S \cup \mathcal{V}^T$ and $\mathcal{V}^S \cap \mathcal{V}^T = \varnothing$. 
\\\\ 
\textbf{Problem statement:} Given the partially available ratings for the source and target domains $r_{uj}, \forall (u,j) \in \Omega^S \cup \Omega^T$, our aim is to predict target domain ratings $r_{uj}$, $\forall (u,j) \not \in \Omega^T$.
\section{Proposed Model}
\label{sec:neucdcf}
In this section, we explain our proposed model -- NeuCDCF in detail. First we explain the individual networks, GCMF and SED, followed by that we show how these two are fused to get NeuCDCF. 
\begin{figure*}[h]
    \centering
    \includegraphics[width=0.8\textwidth]{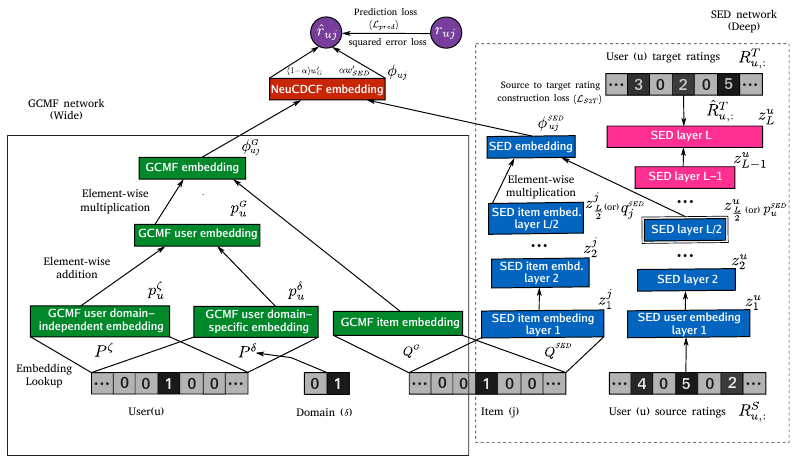} 
\vspace{-.3cm}
    \caption{The architecture of our proposed NeuCDCF Model. GCMF learns two embeddings for shared users -- $p_u^{\zeta}$ and $p_u^{\delta}$. SED learns non-linear embeddings for users and items. In SED, embeddings of shared users are obtained through the network that constructs target domain ratings from source domain ratings. Further, loss $\mathcal{L}_{pred}$ is responsible for rating prediction while loss $\mathcal{L}_{S2T}$ is responsible for source to target rating construction.}
    \label{fig:neucdcf}
\vspace{-.2cm}
\end{figure*}
\subsection{Generalized Collective Matrix Factorization (GCMF)}
\label{subsec:gmf}

GCMF network is shown inside solid-lined box in Figure \ref{fig:neucdcf}. Let $P^{\delta} \in \mathbb{R}^{m \times k}$ and $P^{\zeta} \in \mathbb{R}^{m \times k}$ respectively be domain-specific and domain-independent user embedding matrices, $Q^{\scaleto{G}{3.5pt}} \in \mathbb{R}^{n \times k} $ be an item embedding matrix and $k$ be the embedding dimension. Here, $Q^{\scaleto{G}{3.5pt}}$ encodes items from all the domains collectively such that embeddings for the items can be obtained independent of their domains. Note that, we have a single $P^{\zeta}$ shared across the domains and separate $P^{\delta}$ for source and target domains. Having such separate user embedding matrix $P^{\delta}$ for each domain $\delta$ helps in handling diversity across domains. We define domain-specific embedding ($p^{\delta}_u$) and domain-independent embedding ($p^{\zeta}_u$) for user $u$ as follows:
\begin{equation}
p_u^{\delta} = P^{\delta} x_{u}, ~~ p_u^{\zeta} = P^{\zeta} x_{u} \label{eqn:ui_embed},
\end{equation}
where $p_u^{\delta}$ is obtained from the corresponding embedding matrix $P^{\delta}$ for the domain $\delta$ and $x_u$ and $y_j$ are one-hot encoding of user $u$ and item $j$, respectively.
Now, we find the user and item embeddings $p_u^{\scaleto{G}{3.5pt}}$ and $q_j^{\scaleto{G}{3.5pt}}$ for $(u,j)$ as follows:
\begin{equation}
p_u^{\scaleto{G}{3.5pt}} = p_u^{\zeta} +  p_u^{\delta}, ~~q_j^{\scaleto{G}{3.5pt}} = Q^{\scaleto{G}{3.5pt}} y_{j}. \label{eqn:p_G_rep}
\end{equation}
Here, we use two embeddings $p^{\delta}_u$ and $p^{\zeta}_u$ to define $p^{\scaleto{G}{3.5pt}}_u$ in order to learn the domain-specific features separately from domain independent features. Further, we use the embedding $q_j^{\scaleto{G}{3.5pt}}$ to represent the item $j$. During training, $p^{\zeta}_u$ gets updated irrespective of its interactions with the items from different domains, whereas $p^{\delta}_u$ gets updated only with respect to its specific domain. 

We have different choices in defining $p_u^{\scaleto{G}{3.5pt}}$ using $p^{\delta}_u$ and $p^{\zeta}_u$ other than the one given in Equation (\ref{eqn:p_G_rep}). For example, instead of adding $p^{\delta}_u$ and $p^{\zeta}_u$, one can concatenate them or use the element-wise product ($p^{\delta}_u \odot p^{\zeta}_u$) to obtain $p_u^{\scaleto{G}{3.5pt}}$, and make the appropriate modifications in item embedding vector $q_j^{\scaleto{G}{3.5pt}}$. However, intuitively we can think $p^{\delta}_u$ as an offset to $p^{\zeta}_u$ to account for the domain-specific preferences (e.g. embeddings $p_1^{movie},p_1^{music}$ and $p_1^{book}$ for movie, music and book in Figure \ref{fig:user_rep}) that provide the resultant user representation $p_u^{\scaleto{G}{3.5pt}}$ according to the domain. This also encourages the model to flexibly learn the item embeddings suitable for the domains.

Further, we obtain the representation for the user-item interaction $(u,j)$ and the rating prediction ($\hat{r}_{uj}^{\scaleto{G}{3.5pt}}$) as follows:
\begin{equation}
\phi^{\scaleto{G}{3.5pt}}_{uj} = p_u^{\scaleto{G}{3.5pt}} \odot q_j^{\scaleto{G}{3.5pt}}, ~~\hat{r}_{uj}^{\scaleto{G}{3.5pt}} = a(w_{\scaleto{G}{3.5pt}}' \phi^{\scaleto{G}{3.5pt}}_{uj}),
\label{eqn:gcmf_rep}
\end{equation}
where $\phi^{\scaleto{G}{3.5pt}}_{uj}$ denotes the representation for the user-item pair $(u,j)$, $w_{\scaleto{G}{3.5pt}}$ is a weight vector, and $a(\cdot)$ denotes the activation function with an appropriate scaling according to the maximum rating value $\gamma_{max}$ in the rating matrix. 
Here, we need to scale the output according to the activation function because, for instance, using sigmoid ($\sigma(\cdot)$) as an activation function provides the output value in the range $(0,1)$. So, we use $\gamma_{max} . \sigma(w_{\scaleto{G}{3.5pt}}' \phi^{\scaleto{G}{3.5pt}}_{uj})$ to match the output range to the rating value range in the dataset.

Using definition for  $\hat{r}_{uj}^{\scaleto{G}{3.5pt}}$ as in Equation (\ref{eqn:gcmf_rep}), letting $\Theta = \{P^{\delta},P^{\zeta}$,$Q^{\scaleto{G}{3.5pt}}$, $w_{\scaleto{G}{3.5pt}}\}$ for $\delta \in \{0,1\}$ (for source and target respectively) and leveraging ratings from both domains lead us to the loss function for GCMF:
\begin{equation}
\mathcal{L}_{GCMF} (\Theta) = \frac{1}{|\Omega^S \cup \Omega^T |} \sum_{(u,j) \in \Omega^S \cup \Omega^T} (\hat{r}_{uj}^{\scaleto{G}{3.5pt}} - r_{uj})^2.
\label{eqn:gcmf_loss}
\end{equation}

GCMF is a wide network and it has an ability to generalize many well-known matrix factorization models such as PMF~\cite{mnih2008probabilistic}, CMF~\cite{singh2008relational} and GMF~\cite{he2017neural}. That is, we obtain CMF by assigning $w_{\scaleto{G}{3.5pt}}$ to the vector of all ones and setting $p^{\delta} = 0$ ; we get GMF by setting $p^{\delta} = 0$ and using only single domain ratings; and we obtain PMF by assigning $w_{\scaleto{G}{3.5pt}}$ to the vector of all ones, setting $p^{\delta} = 0$ and using only single domain ratings. 

\subsection{Stacked Encoder-Decoder (SED)}
\label{subsec:sed}

Inspired by the ability of constructing target language sentences from source language in neural machine translation \cite{cho2014learning,bahdanau2014neural,cho2014properties}, we adopt encoder-decoder network for cross-domain recommendation tasks. The proposed SED is illustrated in Figure \ref{fig:neucdcf} inside dashed-lined box. For the sake of brevity, we include the feed-forward network that provides the item embedding ($q_j^{\scaleto{SED}{2.75pt}}$) from the one-hot encoding vector ($y_j$), and the SED embedding ($\phi_{uj}^{\scaleto{SED}{2.75pt}}$) layer as part of SED.

As we discussed earlier, it is essential to capture the non-linear relationships that exist among entities within and across domains. The importance of such representations has also been well argued in \cite{he2017neural,rendle2010factorization,liu2015non}. While constructing target domain ratings from source domain, non-linear embeddings of shared users, and items are obtained as illustrated in Figure~\ref{fig:neucdcf}.

Encoding part of the SED is a feed-forward network that takes partially observed ratings for users from the source domain, $R_{u,:}^S \in \mathbb{R}^{n_S}, \forall u \in 
\mathcal{U}$, projects them into a low dimensional ($z_{\frac{L}{2}}^{u} \in \mathbb{R}^k$) embedding space as follows:
\begin{equation}
 z^u_{\frac{L}{2}} = a(W^U_\frac{L}{2} (a(W^U_{\frac{L}{2}-1} (\dots a(W^U_1 R_{u,:}^S + b^U_1) \dots) + b^U_{\frac{L}{2}-1})) + b^U_\frac{L}{2}),  \label{eqn:sed_encoder}
\end{equation}
where $z_l^u,W_l^U$ and $b_l^U$ denote user representation, weight matrix and bias at layer $l$, respectively. At layer $\frac{L}{2}$, the embedding vector $z^u_{\frac{L}{2}}$ acts as a low-dimensional compact representation ($p_u^{\scaleto{SED}{2.75pt}}$) for user $u$. Here, $p_u^{\scaleto{SED}{2.75pt}}$ contains the transferable user behavior component for user $u$ from source to target domain.

Further, decoder is also a feed-forward network that takes this low dimensional embedding ($z_{\frac{L}{2}}^{u}$) and constructs the user ratings for the target domain ($\hat{R}_{u,:}^T \in \mathbb{R}^{n_T}$) as follows:
\begin{align}
z^u_L = a(W^U_L (a(W^U_{L-1} (\dots a(W^U_1 z_{\frac{L}{2}}^{u} + b^U_{\frac{L}{2}}) \dots) + b^U_{L-1})) + b^U_L),  \label{eqn:sed_decoder}
\end{align}
where, $z^u_{L}$ is treated as the constructed ratings ($\hat{R}_{u,:}^T$) of user $u$ for the target domain. Hence, we have the following loss function for source to target domain rating construction:
\begin{equation}
\mathcal{L}_{S2T}(\Theta) = \frac{1}{|\Omega^T|} \sum_{u \in \mathcal{U}} \|(z^u_{L} - R_{u,:}^T) \odot \mathbbm{1}(R_{u,:}^T>0)\|_2^2,
\label{eqn:sed_loss}
\end{equation}
where $\Theta$ is a set of parameters that contains all the weight matrices and bias vectors, and $\mathbbm{1}(\cdot)$ denotes indicator function. 

In addition to this, we get the item representation from SED as follows:
\begin{align}
\begin{split}
 q_j^{\scaleto{SED}{3.5pt}} = z^j_{\frac{L}{2}}, \text{~~where~~} z^j_{\frac{L}{2}} = a(W^V_{\frac{L}{2}} (a(W^V_{\frac{L}{2}-1} (\dots a(W^V_1 Q^{\scaleto{SED}{3.5pt}} y_j \\+ b^V_1) \dots) + b^V_{\frac{L}{2}-1})) + b^V_{\frac{L}{2}}).
\end{split}
\label{eqn:sed_item_rep}
\end{align} 
Here $Q^{\scaleto{SED}{3.5pt}} \in \mathbb{R}^{n \times k}$ denotes the item embedding matrix and $q_j^{\scaleto{SED}{3.5pt}}$ represents the embedding vector for the item $j$ in SED. $q_j^{\scaleto{SED}{3.5pt}}$ is further used along with $p_u^{\scaleto{SED}{3.5pt}}$ to predict the target domain ratings as shown in Figure \ref{fig:neucdcf}.

Note that, one can use the same item embedding matrix $Q^{\scaleto{G}{3.5pt}}$ in (Equation (\ref{eqn:sed_item_rep})) the SED network instead of $Q^{\scaleto{SED}{3.5pt}}$. Nevertheless, we use a separate embedding matrix $Q^{\scaleto{SED}{3.5pt}}$ for the following reasons: using the same $Q^{\scaleto{G}{3.5pt}}$ here restricts the flexibility in using different number of activation functions at the SED layer $L/2$; we want to obtain low dimensional deep representations from SED and wide representations from GCMF, and using $q_j^{\scaleto{G}{3.5pt}}$ (GCMF item vector in Figure \ref{fig:neucdcf}) in both places for item $j$ may deteriorate the resultant representations. 

Here, $p_u^{\scaleto{SED}{3.5pt}}$ ($z^u_{\frac{L}{2}}$) acts as a component that transfers the user behavior to the target domain. In addition, $p_u^{\scaleto{SED}{3.5pt}}$ is particularly helpful to learn the representations for the users when they have less or no ratings in the target domain. Further, stacked layers help in extracting the complicated relationships present between entities.

Similar to GCMF, we learn the representations $\phi^{\scaleto{SED}{3.5pt}}_{uj}$ and predict rating $\hat{r}_{uj}^{\scaleto{SED}{3.5pt}}$ for the user-item pair $(u,j)$ as follows:
\begin{equation}
 \hat{r}_{uj}^{\scaleto{SED}{3.5pt}} = a(w_{\scaleto{SED}{3.5pt}}' \phi^{\scaleto{SED}{3.5pt}}_{uj}), ~ \text{~where~}~\phi^{\scaleto{SED}{3.5pt}}_{uj} = p_u^{\scaleto{SED}{3.5pt}} \odot q_j^{\scaleto{SED}{3.5pt}}.
\label{eqn:sed_rep}
\end{equation}
Here $w_{\scaleto{SED}{3.5pt}}$ denotes the weight vector. The loss function of SED ($\mathcal{L}_{\scaleto{SED}{3.5pt}}$) contains two terms: rating prediction loss ($\mathcal{L}_{pred}$) and source to target rating construction loss ($\mathcal{L}_{S2T}$). Let $\Omega = \Omega^S \cup \Omega^T$. The loss function is given as follows: 
\begin{align}
\begin{split}
\mathcal{L}_{\scaleto{SED}{3.5pt}}(\Theta) = \frac{1}{|\Omega|} \sum_{(u,j) \in \Omega} (\hat{r}_{uj}^{\scaleto{SED}{3.5pt}} - r_{uj})^2 \\
+ \frac{1}{|\Omega^T|} \sum_{u \in \mathcal{U}} \|(\hat{R}_{u,:}^T - R_{u,:}^T) \odot \mathbbm{1}(R_{u,:}^T>0)\|_2^2.
\end{split}
\label{eqn:sed_loss}
\end{align}

\subsection{Fusion of GCMF and SED}
\label{subsec:neucdcf}
We fuse GCMF and SED network to get the best of both wide and deep representations. The embedding for NeuCDCF, and the rating prediction ($\hat{r}_{uj}$) is obtained as follows: 
\begin{align}
\begin{split}
\hat{r}_{uj} = a(w' \phi_{uj}),
~\text{~where~}
\phi_{uj} = \begin{bmatrix}
         	\phi^{\scaleto{G}{3.5pt}}_{uj}
      	\\ 	\phi^{\scaleto{SED}{3.5pt}}_{uj}
        \end{bmatrix},
 ~~~ w = \begin{bmatrix}
         	(1-\alpha) ~~ w_{\scaleto{G}{3.5pt}}
         \\ \alpha \text{~~~~~~} w_{\scaleto{SED}{3.5pt}}
        \end{bmatrix} \label{eqn:neucdcf_rep}.
\end{split}
\end{align}
Here $\alpha \in [0,1]$ is a trade-off parameter between GCMF and SED networks and it is tuned using cross-validation. When $\alpha$ is explicitly set to $0$ and $1$, NeuCDCF considers only GCMF network and SED network, respectively. 

The objective function for NeuCDCF is the sum of three individual loss functions:
\begin{align}
\begin{split}
\mathcal{L}(\Theta) = \mathcal{L}_{pred} + \mathcal{L}_{S2T} + \mathcal{L}_{reg}, \\
\text{where} ~~~
\mathcal{L}_{pred} = \frac{1}{|\Omega^S \cup \Omega^T |} \sum_{(u,j) \in \Omega^S \cup \Omega^T} (\hat{r}_{uj} - r_{uj})^2, \\
\mathcal{L}_{S2T} = \frac{1}{|\Omega^T|} \sum_{u \in \mathcal{U}} \|(\hat{R}_{u,:}^T - R_{u,:}^T) \odot \mathbbm{1}(R_{u,:}^T>0)\|_2^2~,\\
\text{and~~} 
\mathcal{L}_{reg} = \mathcal{R}(\Theta). \label{eqn:neucdcf_loss}
\end{split}
\end{align}
Here $\mathcal{L}_{pred}$ is the prediction loss from both the source and target domains together, $\mathcal{L}_{S2T}$ denotes the source to target rating construction loss between predicted ratings $\hat{R}_{u,:}^T$ and the available true ratings ${R}_{u,:}^T$ for all the users $u \in \mathcal{U}$, and $\mathcal{L}_{reg}$ is the regularization loss derived using all the parameters in the model. Here, $\mathcal{L}_{S2T}$ also acts as a regularizer that helps in properly learning $p_u^{\scaleto{SED}{3.5pt}}$ and $\phi^{\scaleto{SED}{3.5pt}}_{uj}$. 

Note that the proposed architecture can easily be adapted to implicit rating settings where only binary ratings available (e.g., clicks or likes) by changing the loss function from squared error loss to cross-entropy \cite{he2017neural} or pairwise loss \cite{rendle2009bpr} with appropriate negative sampling strategy \cite{mikolov2013distributed}.

Since the datasets we experiment with contain real-valued ratings, the choice of squared error loss is natural. In the next section, we demonstrate the effectiveness of our model on various real-world datasets.
\section{Experiments}
\label{sec:exp}
We conduct our experiments to address the following questions:
\begin{description}
    \item{\textbf{RQ1}} How do GCMF, SED and NeuCDCF perform under the different sparsity and cold-start settings?
    \item{\textbf{RQ2}} Does fusing wide and deep representations boost the performance?
    \item{\textbf{RQ3}} Is adding domain-specific component helpful in improving the performance of GCMF?
    \item{\textbf{RQ4}} Does capturing non-linear relationship by SED provide any benefits?
\end{description}
We first present the experimental settings after that we answer the above questions.
\begin{table}[h]
\begin{center}
\scalebox{.8}{
\begin{tabular}{c|c|c|c|c|c|c}
    \hline
     \multicolumn{2}{c|}{Dataset~~~}& \# users & \# items & \# ratings & ~density~ & role \\ \hline   
    \multirow{4}{*}{Amazon}    
     &Movie (M) & 3677 & 7549 & 222095 & 0.80 \% & source\\
     &Book  (B) & 3677 & 9919 & 183763 & 0.50 \% & target
     \\\cline{2-7}
     &Movie (M)  & 2118 & 5099 & 223383 & 2.07 \% & source\\
     &Music (Mu) & 2118 & 4538 &  91766 & 0.95 \% & target\\
     \hline\hline
     \multirow{4}{*}{Douban} 
     &Movie (M) & 5674 & 11676 & 1863139 & 2.81 \% & source\\
     &Book  (B) & 5674 & 6626  &  432642 & 1.15 \% & target
     \\\cline{2-7}    
     &Music (Mu) & 3628 & 8118 & 554075 & 1.88 \% & source\\
     &Book  (B) & 3628 & 4893 & 293488 & 1.65 \% & target\\
     \hline     
\end{tabular}
}
\end{center}
\caption{Statistics of the Amazon and Douban datasets}
\vspace{-.3cm}
\label{tab:dataset_stat}
\end{table}
\vspace{-0.5cm}
\subsection{Experimental Setup}
\subsubsection{Datasets.} We adopt four widely used datasets: two from Amazon \cite{he2016ups} -- an e-commerce product recommendation system, and two from Douban \cite{zhong2014user} -- a Chinese online community platform for movies, books and music recommendations. All the datasets are publicly available\footnote{http://jmcauley.ucsd.edu/data/amazon (we rename CDs-and-Vinyl as Music)}\footnote{https://sites.google.com/site/erhengzhong/datasets} and have the ratings from multiple domains. We do the following preprocessing similar to the one done in \cite{man2017cross,gao2013cross,xue2017deep}: we retain (1) only those ratings associated with the users who are common between both the domains and (2) the users and items that have at least 10 and 20 ratings (different thresholds are used here to have diversity in density across datasets) for Amazon and Douban datasets, respectively. Statistics of the datasets are given in Table \ref{tab:dataset_stat}.
\begin{table*}[h]
\begin{center}
\scalebox{.74}{
\begin{tabular}{c|c|c|c|c|c|c||c|c|c}
    \hline
    \multicolumn{7}{c||}{Sparse} & \multicolumn{3}{c}{Cold-start}  \\ \hline
    Dataset & Model & $K$=0\% & $K$=20\% &  $K$=40\% & $K$=60\% & $K$=80\% & $K$=40\% & $K$=60\% & $K$=80\% \\ \hline
     
                   & PMF~\cite{mnih2008probabilistic}      & 0.7481 $\pm$ 0.0042 & 0.7828 $\pm$ 0.0061 & 0.8449 $\pm$ 0.0178 & 0.9640 $\pm$ 0.0035 & 1.2223 $\pm$ 0.0145                                                                                          & 1.1063 $\pm$ 0.0501 & 1.1476 $\pm$ 0.0250 & 1.5210 $\pm$ 0.0316\\
                   & GMF~\cite{he2017neural}      & 0.7029 $\pm$ 0.0110 & 0.7108 $\pm$ 0.0056 & 0.7254 $\pm$ 0.0078 & 0.7805 $\pm$ 0.0151 & 0.8878 $\pm$ 0.0207                                                                               & 0.9040 $\pm$ 0.0704 & 1.0721 $\pm$ 0.0791 & 1.0735 $\pm$ 0.0250\\ \cline{2-10}
		   & GMF-CD~\cite{he2017neural}  & 0.6961 $\pm$ 0.0012 & 0.7085 $\pm$ 0.0056 & 0.7186 $\pm$ 0.0018 & 0.7552 $\pm$ 0.0236 & 0.8009 $\pm$ 0.0325                                                                                           & 0.8086 $\pm$ 0.0296 & 0.8220 $\pm$ 0.0158 & 0.8792 $\pm$ 0.0144\\
    Amazon    & CMF~\cite{singh2008relational}      & 0.7242 $\pm$ 0.0022 & 0.7434 $\pm$ 0.0044 & 0.7597 $\pm$ 0.0019 & 0.7919 $\pm$ 0.0039 & 0.8652 $\pm$ 0.0032                                                                                          & 0.8617 $\pm$ 0.0327 & 0.8734 $\pm$ 0.0126 & 0.9362 $\pm$ 0.0129\\
        (M-B)           & MV-DNN~\cite{elkahky2015multi}    & 0.7048 $\pm$ 0.0024 & 0.7160 $\pm$ 0.0047 & 0.7192 $\pm$ 0.0035 & 0.7255 $\pm$ 0.0019 & 0.7624 $\pm$ 0.0019                                                                              & 0.8052 $\pm$ 0.0293 & 0.8135 $\pm$ 0.0167 & 0.8347 $\pm$ 0.0243\\ \cline{2-10}
                   & GCMF (ours)    & 0.6705 $\pm$ 0.0019 & 0.6862 $\pm$ 0.0052 & 0.7023 $\pm$ 0.0025 & 0.7254 $\pm$ 0.0023 & 0.7873 $\pm$ 0.0030                                                                                          & 0.8086 $\pm$ 0.0296 & 0.8220 $\pm$ 0.0158 & 0.8792 $\pm$ 0.0144\\
                   & SED  (ours)     & 0.6880 $\pm$ 0.0033 & 0.7024 $\pm$ 0.0060 & 0.7094 $\pm$ 0.0030 & 0.7194 $\pm$ 0.0023 & 0.7428 $\pm$ 0.0024                                                                                          & 0.7925 $\pm$ 0.0301 & 0.8002 $\pm$ 0.0157 & 0.8169 $\pm$ 0.0118\\
                   & NeuCDCF (ours)  & \textbf{0.6640} $\pm$ \textbf{0.0018} & \textbf{0.6786} $\pm$ \textbf{0.0052} & \textbf{0.6852} $\pm$ \textbf{0.0025} & \textbf{0.6916} $\pm$ \textbf{0.0010} & \textbf{0.7291} $\pm$ \textbf{0.0060}& \textbf{0.7830} $\pm$ \textbf{0.0304} & \textbf{0.7791} $\pm$ \textbf{0.0165} & \textbf{0.8015} $\pm$ \textbf{0.0157}\\ \hline \hline                    
                   & PMF~\cite{mnih2008probabilistic}      & 0.7135 $\pm$ 0.0105 & 0.7366 $\pm$ 0.0050 & 0.7926 $\pm$ 0.0118 & 0.9324 $\pm$ 0.0174 & 1.2080 $\pm$ 0.0186                                                                                          & 1.0552 $\pm$ 0.0315 & 1.1734 $\pm$ 0.0577 & 1.5235 $\pm$ 0.0330\\
                   & GMF~\cite{he2017neural}      & 0.6583 $\pm$ 0.0056 & 0.6731 $\pm$ 0.0090 & 0.7008 $\pm$ 0.0083 & 0.7268 $\pm$ 0.0194 & 0.8331 $\pm$ 0.0046                                                                               & 0.8425 $\pm$ 0.0687 & 0.8490 $\pm$ 0.0345 & 1.0013 $\pm$ 0.0155\\ \cline{2-10}
		   & GMF-CD~\cite{he2017neural}  & 0.6628 $\pm$ 0.0025 & 0.6666 $\pm$ 0.0047 & 0.6883 $\pm$ 0.0175 & 0.6997 $\pm$ 0.0061 & 0.7566 $\pm$ 0.0277                                                                                           & 0.7537 $\pm$ 0.0262 & 0.7600 $\pm$ 0.0252 & 0.8295 $\pm$ 0.0147 \\
    Amazon   & CMF~\cite{singh2008relational}      & 0.6923 $\pm$ 0.0041 & 0.7071 $\pm$ 0.0013 & 0.7298 $\pm$ 0.0103 & 0.7594 $\pm$ 0.0071 & 0.8279 $\pm$ 0.0046                                                                                          & 0.8359 $\pm$ 0.0320 & 0.8245 $\pm$ 0.0139 & 0.8746 $\pm$ 0.0194\\
        (M-Mu)           & MV-DNN~\cite{elkahky2015multi}    & 0.6843 $\pm$ 0.0017 & 0.6827 $\pm$ 0.0046 & 0.6898 $\pm$ 0.0077 & 0.6905 $\pm$ 0.0037 & 0.7018 $\pm$ 0.0031                                                                              & 0.7320 $\pm$ 0.0253 & \textbf{0.7075} $\pm$ \textbf{0.0313} & \textbf{0.7241} $\pm$ \textbf{0.0157}\\ \cline{2-10}
                   & GCMF  (ours)    & 0.6173 $\pm$ 0.0053 & 0.6261 $\pm$ 0.0042 & 0.6527 $\pm$ 0.0111 & 0.6809 $\pm$ 0.0040 & 0.7436 $\pm$ 0.0038                                                                                          & 0.7537 $\pm$ 0.0262 & 0.7600 $\pm$ 0.0252 & 0.8295 $\pm$ 0.0147\\
                   & SED  (ours)     & 0.6406 $\pm$ 0.0032 & 0.6430 $\pm$ 0.0040 & 0.6597 $\pm$ 0.0114 & 0.6770 $\pm$ 0.0039 & 0.7006 $\pm$ 0.0044                                                                                          & 0.7281 $\pm$ 0.0314 & 0.7176 $\pm$ 0.0265 & 0.7437 $\pm$ 0.0156\\
                   & NeuCDCF  (ours) & \textbf{0.6095} $\pm$ \textbf{0.0043} & \textbf{0.6170} $\pm$ \textbf{0.0044} & \textbf{0.6318} $\pm$ \textbf{0.0101} & \textbf{0.6535} $\pm$ \textbf{0.0050} & \textbf{0.6937} $\pm$ \textbf{0.0061}& \textbf{0.7144} $\pm$ \textbf{0.0232} & 0.7134 $\pm$ 0.0267 & 0.7390 $\pm$ 0.0161\\ \hline \hline                                                        
                   & PMF~\cite{mnih2008probabilistic}     & 0.5733 $\pm$ 0.0019 & 0.5811 $\pm$ 0.0018 & 0.5891 $\pm$ 0.0020 & 0.6136 $\pm$ 0.0016 & 0.7141 $\pm$ 0.0066                                                                                           & 0.8654 $\pm$ 0.0593 & 0.7527 $\pm$ 0.0199 & 0.8275 $\pm$ 0.0135\\
                   & GMF~\cite{he2017neural}     & 0.5730 $\pm$ 0.0016 & 0.5747 $\pm$ 0.0017 & 0.5781 $\pm$ 0.0033 & 0.5971 $\pm$ 0.0071 & 0.6133 $\pm$ 0.0072                                                                                & 0.6205 $\pm$ 0.0530 & 0.7290 $\pm$ 0.0740 & 0.6625 $\pm$ 0.0371\\ \cline{2-10}
                   & GMF-CD~\cite{he2017neural} & 0.5803 $\pm$ 0.0026 & 0.5836 $\pm$ 0.0029 & 0.5879 $\pm$ 0.0031 & 0.6002 $\pm$ 0.0089 & 0.6117 $\pm$ 0.0078                                                                                            & 0.5495 $\pm$ 0.0625 & 0.5983 $\pm$ 0.0172 & 0.6220 $\pm$ 0.0037\\
    Douban    & CMF~\cite{singh2008relational}     & 0.5771 $\pm$ 0.0006 & 0.5821 $\pm$ 0.0014 & 0.5862 $\pm$ 0.0012 & 0.5979 $\pm$ 0.0025 & 0.6188 $\pm$ 0.0016                                                                                           & 0.5490 $\pm$ 0.0711 & 0.5998 $\pm$ 0.0156 & 0.6171 $\pm$ 0.0069\\
        (M-B)           & MV-DNN~\cite{elkahky2015multi}   & 0.5956 $\pm$ 0.0005 & 0.6009 $\pm$ 0.0019 & 0.6039 $\pm$ 0.0011 & 0.6127 $\pm$ 0.0023 & 0.6224 $\pm$ 0.0015                                                                               & 0.5624 $\pm$ 0.0807 & 0.6110 $\pm$ 0.0165 & 0.6206 $\pm$ 0.0113\\ \cline{2-10}
                   & GCMF (ours)    & 0.5608 $\pm$ 0.0009 & 0.5664 $\pm$ 0.0023 & 0.5709 $\pm$ 0.0016 & 0.5849 $\pm$ 0.0023 & 0.6118 $\pm$ 0.0018                                                                                           & 0.5495 $\pm$ 0.0625 & 0.5983 $\pm$ 0.0172 & 0.6220 $\pm$ 0.0037\\
                   & SED (ours)     & 0.5822 $\pm$ 0.0003 & 0.5865 $\pm$ 0.0021 & 0.5905 $\pm$ 0.0018 & 0.6011 $\pm$ 0.0029 & 0.6124 $\pm$ 0.0022                                                                                           & 0.5462 $\pm$ 0.0671 & 0.5983 $\pm$ 0.0159 & 0.6105 $\pm$ 0.0050\\
		   & NeuCDCF (ours) & \textbf{0.5603} $\pm$ \textbf{0.0009} & \textbf{0.5647} $\pm$ \textbf{0.0021} & \textbf{0.5704} $\pm$ \textbf{0.0014} & \textbf{0.5800} $\pm$ \textbf{0.0023} & \textbf{0.5957} $\pm$ \textbf{0.0019} & \textbf{0.5372} $\pm$ \textbf{0.0699} & \textbf{0.5911} $\pm$ \textbf{0.0157} & \textbf{0.6031} $\pm$ \textbf{0.0033}\\ \hline \hline
                   & PMF~\cite{mnih2008probabilistic}     & 0.5750 $\pm$ 0.0022 & 0.5800 $\pm$ 0.0016 & 0.5894 $\pm$ 0.0033 & 0.6146 $\pm$ 0.0037 & 0.7319 $\pm$ 0.0099                                                                                           & 0.9201 $\pm$ 0.0951 & 0.7629 $\pm$ 0.0237 & 0.8451 $\pm$ 0.0161\\
                   & GMF~\cite{he2017neural}     & 0.5745 $\pm$ 0.0033 & 0.5768 $\pm$ 0.0036 & 0.5765 $\pm$ 0.0018 & 0.5900 $\pm$ 0.0051 & 0.6241 $\pm$ 0.0153                                                                                & 0.5489 $\pm$ 0.0594 & 0.6873 $\pm$ 0.0731 & 0.6964 $\pm$ 0.0634\\ \cline{2-10}
                   & GMF-CD~\cite{he2017neural} & 0.5825 $\pm$ 0.0023 & 0.5847 $\pm$ 0.0040 & 0.5883 $\pm$ 0.0040 & 0.5962 $\pm$ 0.0067 & 0.6137 $\pm$ 0.0037                                                                                             & 0.5220 $\pm$ 0.0717 & 0.6129 $\pm$ 0.0212 & 0.6205 $\pm$ 0.0047\\ 
     Douban    & CMF~\cite{singh2008relational}     & 0.5827 $\pm$ 0.0017 & 0.5881 $\pm$ 0.0015 & 0.5933 $\pm$ 0.0024 & 0.6035 $\pm$ 0.0017 & 0.6231 $\pm$ 0.0019                                                                                           & 0.5875 $\pm$ 0.0543 & 0.6081 $\pm$ 0.0208 & 0.6194 $\pm$ 0.0060\\
          (Mu-B)         & MV-DNN~\cite{elkahky2015multi}   & 0.5892 $\pm$ 0.0015 & 0.5918 $\pm$ 0.0013 & 0.5946 $\pm$ 0.0016 & 0.6039 $\pm$ 0.0022 & 0.6180 $\pm$ 0.0022                                                                               & 0.5485 $\pm$ 0.0367 & 0.6131 $\pm$ 0.0198 & 0.6089 $\pm$ 0.0049\\ \cline{2-10}
                   & GCMF  (ours)   & 0.5675 $\pm$ 0.0012 & 0.5707 $\pm$ 0.0013 & 0.5768 $\pm$ 0.0018 & 0.5905 $\pm$ 0.0020 & 0.6187 $\pm$ 0.0021                                                                                           & 0.5220 $\pm$ 0.0717 & 0.6129 $\pm$ 0.0212 & 0.6205 $\pm$ 0.0047\\
                   & SED  (ours)    & 0.5769 $\pm$ 0.0024 & 0.5782 $\pm$ 0.0019 & 0.5839 $\pm$ 0.0020 & 0.5934 $\pm$ 0.0023 & 0.6090 $\pm$ 0.0025                                                                                           & 0.5423 $\pm$ 0.0380 & 0.6011 $\pm$ 0.0239 & 0.5999 $\pm$ 0.0046\\
                   & NeuCDCF (ours) & \textbf{0.5625} $\pm$ \textbf{0.0015} & \textbf{0.5646} $\pm$ \textbf{0.0013} & \textbf{0.5688} $\pm$ \textbf{0.0017} & \textbf{0.5794} $\pm$ \textbf{0.0021} & \textbf{0.5954} $\pm$ \textbf{0.0024} & \textbf{0.5200} $\pm$ \textbf{0.0482} & \textbf{0.5963} $\pm$ \textbf{0.0246} & \textbf{0.5954} $\pm$ \textbf{0.0046}\\ \hline \hline
\end{tabular}
}
\end{center}
\caption{\textbf{Sparsity and Cold-start} -- Performance of different models on four real-world datasets for different sparsity levels. Here, $K$\% indicates the percentage of the target domain ratings removed from the training set. In cold-start setting we set $\mathbf{p_u^{\delta}}$ to vector of zeros, hence, GCMF reduces to GMF-CD (refer section \ref{subsubsec:res_dis} for more details). The bold-faced values indicate the best performance for every dataset. We conduct paired $t$-test and the improvements are statistically significant with $p<0.01$.}
    \vspace{-.2cm}
\label{tab:sparsity}
\end{table*}

\begin{table*}[h]
\begin{center}
\scalebox{.85}{
\begin{tabular}{c|c|c|c|c|c|c}
    \hline
    \multicolumn{7}{c}{Full-cold-start setting}  \\ \hline
    Dataset & Model & $K$=10 & $K$=20 &  $K$=30 & $K$=40 & $K$=50 \\ \hline
                   & CMF~\cite{singh2008relational}       & 0.7870 $\pm$ 0.0079 & 0.7856 $\pm$ 0.0115 & 0.8017 $\pm$ 0.0286 & 0.8214 $\pm$ 0.0170 & 0.8590 $\pm$ 0.0327    \\                   
                   & EMCDR~\cite{man2017cross}    & 0.7340 $\pm$ 0.0042 & 0.7324 $\pm$ 0.0089 & 0.7436 $\pm$ 0.0195 & 0.7831 $\pm$ 0.0167 & 0.8067 $\pm$ 0.0211\\
  Amazon (M-B)     & MV-DNN~\cite{elkahky2015multi}   & 0.7401 $\pm$ 0.0038 & 0.7366 $\pm$ 0.0086 & 0.7328 $\pm$ 0.0161 & 0.7415 $\pm$ 0.0047 & 0.7483 $\pm$ 0.0075 \\ \cline{2-7}
                   & GCMF  (ours)    & 0.7233 $\pm$ 0.0076 & 0.7208 $\pm$ 0.0080 & 0.7226 $\pm$ 0.0177 & 0.7371 $\pm$ 0.0081 & 0.7590 $\pm$ 0.0034 \\
                   & SED   (ours)     & 0.7181 $\pm$ 0.0040 & 0.7108 $\pm$ 0.0062 & 0.7115 $\pm$ 0.0151 & 0.7178 $\pm$ 0.0060 & 0.7277 $\pm$ 0.0044   \\
                   & NeuCDCF  (ours)   & \textbf{0.7096} $\pm$ \textbf{0.0073} & \textbf{0.7012} $\pm$ \textbf{0.0077} & \textbf{0.6983} $\pm$ \textbf{0.0171} & \textbf{0.7091} $\pm$ \textbf{0.0077} & \textbf{0.7215} $\pm$ \textbf{0.0043}   \\ \hline \hline
                   
                   & CMF~\cite{singh2008relational}      & 0.7467 $\pm$ 0.0138 & 0.7413 $\pm$ 0.0123 & 0.7600 $\pm$ 0.0220 & 0.7865 $\pm$ 0.0146 & 0.8348 $\pm$ 0.0309   \\
                   & EMCDR~\cite{man2017cross}    & 0.6781 $\pm$ 0.0147 & 0.6910 $\pm$ 0.0104 & 0.7349 $\pm$ 0.0212 & 0.7682 $\pm$ 0.0172 & 0.7917 $\pm$ 0.0272\\
     Amazon (M-Mu) & MV-DNN~\cite{elkahky2015multi}   & 0.6848 $\pm$ 0.0270 & 0.6923 $\pm$ 0.0170 & 0.6833 $\pm$ 0.0274 & 0.7037 $\pm$ 0.0116 & 0.7178 $\pm$ 0.0135  \\ \cline{2-7}
                   & GCMF  (ours)    & 0.6658 $\pm$ 0.0237 & 0.6766 $\pm$ 0.0154 & 0.6635 $\pm$ 0.0237 & 0.6943 $\pm$ 0.0118 & 0.7280 $\pm$ 0.0104   \\
                   & SED  (ours)     & 0.6523 $\pm$ 0.0281 & 0.6544 $\pm$ 0.0181 & 0.6452 $\pm$ 0.0214 & 0.6693 $\pm$ 0.0115 & 0.6864 $\pm$ 0.0098   \\
                   & NeuCDCF (ours)  & \textbf{0.6418} $\pm$ \textbf{0.0257} & \textbf{0.6471} $\pm$ \textbf{0.0183} & \textbf{0.6384} $\pm$ \textbf{0.0222} & \textbf{0.6650} $\pm$ \textbf{0.0107} & \textbf{0.6856} $\pm$ \textbf{0.0095}   \\ \hline \hline

                   & CMF~\cite{singh2008relational}      & 0.5919 $\pm$ 0.0052 & 0.5987 $\pm$ 0.0062 & 0.5970 $\pm$ 0.0055 & 0.5983 $\pm$ 0.0037 & 0.6004 $\pm$ 0.0012   \\
                   & EMCDR~\cite{man2017cross}    & 0.5942 $\pm$ 0.0069 & 0.5926 $\pm$ 0.0071 & 0.5973 $\pm$ 0.0064 & 0.5996 $\pm$ 0.0032 & 0.6112 $\pm$ 0.0031\\
     Douban (M-B)  & MV-DNN~\cite{elkahky2015multi}   & 0.6004 $\pm$ 0.0071 & 0.6106 $\pm$ 0.0069 & 0.6062 $\pm$ 0.0062 & 0.6068 $\pm$ 0.0025 & 0.6075 $\pm$ 0.0032   \\ \cline{2-7}
                   & GCMF  (ours)    & 0.5852 $\pm$ 0.0070 & 0.5923 $\pm$ 0.0074 & 0.5919 $\pm$ 0.0044 & 0.5930 $\pm$ 0.0021 & 0.5932 $\pm$ 0.0022    \\
                   & SED   (ours)    & 0.5897 $\pm$ 0.0072 & 0.5998 $\pm$ 0.0094 & 0.5955 $\pm$ 0.0066 & 0.5965 $\pm$ 0.0026 & 0.5978 $\pm$ 0.0032    \\
                   & NeuCDCF (ours)  & \textbf{0.5807} $\pm$ \textbf{0.0065} & \textbf{0.5877} $\pm$ \textbf{0.0072} & \textbf{0.5858} $\pm$ \textbf{0.0048} & \textbf{0.5864} $\pm$ \textbf{0.0019} & \textbf{0.5869} $\pm$ \textbf{0.0020}   \\ \hline \hline
                   
                   & CMF~\cite{singh2008relational}      & 0.6063 $\pm$ 0.0112 & 0.5971 $\pm$ 0.0048 & 0.5995 $\pm$ 0.0025 & 0.6000 $\pm$ 0.0019 & 0.5997 $\pm$ 0.0021   \\
                   & EMCDR~\cite{man2017cross}    & 0.6058 $\pm$ 0.0072 & 0.6021 $\pm$ 0.0046 & 0.6083 $\pm$ 0.0034 & 0.6102 $\pm$ 0.0033 & 0.6092 $\pm$ 0.0042\\
     Douban (Mu-B) & MV-DNN~\cite{elkahky2015multi}   & 0.6145 $\pm$ 0.0112 & 0.6053 $\pm$ 0.0033  & 0.6072 $\pm$ 0.0031 & 0.6059 $\pm$ 0.0025 & 0.6035 $\pm$ 0.0035   \\ \cline{2-7}
                   & GCMF (ours)     & 0.6142 $\pm$ 0.0054 & 0.6044 $\pm$ 0.0030 & 0.6115 $\pm$ 0.0037 & 0.6082 $\pm$ 0.0025 & 0.6078 $\pm$ 0.0039    \\
                   & SED (ours)      & 0.6037 $\pm$ 0.0119 & 0.5936 $\pm$ 0.0045 & 0.5962 $\pm$ 0.0033 & 0.5940 $\pm$ 0.0023 & 0.5927 $\pm$ 0.0025   \\
                   & NeuCDCF (ours)  & \textbf{0.5978} $\pm$ \textbf{0.0098} & \textbf{0.5880} $\pm$ \textbf{0.0042} & \textbf{0.5927} $\pm$ \textbf{0.0029} & \textbf{0.5898} $\pm$ \textbf{0.0017} & \textbf{0.5885} $\pm$ \textbf{0.0030}    \\ \hline \hline
\end{tabular}
}
\end{center}
\caption{\textbf{Full-cold-start} -- Performance of different models on four real-world datasets for complete cold-start users on different cold-start levels. Here, $K$\% indicates the percentage of users whose all ratings in the target domain are removed from the training set. The bold-faced values indicate the best performance. We conduct paired $t$-test and the improvements are significant with $p<0.01$.}
\label{tab:full_cold_start}
\vspace{-.2cm}
\end{table*}

\subsubsection{Evaluation procedure.} We randomly split the target domain ratings of the datasets into training (65\%), validation (15\%), and test (20\%) sets. In this process, we make sure that at least one rating for every user and every item is present in the training set. For the CDCF models, we include source domain ratings to the above-mentioned training set. Each experiment is repeated five times for different random splits of the dataset and we report the average test performance and standard deviation with respect to the best validation error as the final results. 
 We evaluate our proposed models (GCMF, SED and NeuCDCF) against baseline models under the following 3 experimental settings:
\begin{enumerate}
\item \textbf{Sparsity:} In this setting, we remove $K$\% of the target domain ratings from the training set. We use different values of $K$ from \{0, 20, 40, 60, 80\} to get different sparsity levels.
\item \textbf{Cold-start:} With the above setting, we report the results obtained for only those users who have less than five ratings in the target domain of the training set. We call them cold-start users. The similar setting was defined and followed in \cite{guo2015trustsvd}. We do not include 0\% and 20\% sparsity levels here as the respective splits have a very few cold-start users.
\item \textbf{Full-cold-start:} We remove all the ratings of $K$\% of users from target domain and call them as full-cold-start users since these users have no ratings at all in the target domain of the training set. We use different values of $K$ from \{10, 20, 30, 40, 50\} to get different full-cold-start levels. This particular setting was followed in \cite{man2017cross}.
\end{enumerate}
\subsubsection{Metric.} We employ Mean Absolute Error (MAE) for performance analysis \cite{cantador2015cross,guo2015trustsvd,li2009can,li2009transfer}. It is defined as follows:
\[MAE =  \frac{1}{|\Omega^{Test}|} \sum_{(u,j)\in \Omega^{Test}} | r_{uj} - \hat{r}_{uj} |, \] 
where $\Omega^{Test}$ is a test set and smaller values of MAE indicate better prediction. We observed similar behaviours in performance in terms  of root mean square error (RMSE)~\cite{cantador2015cross} and due to space limitations the results in RMSE are omitted. 

\subsubsection{Comparison of different models.}
To evaluate the performance of NeuCDCF and its individual networks (GCMF and SED) in CDR setting, we compare our models with the representative models from the following three categories:\\
\textbf{(1) Deep neural network based CDCF models (Deep):}
\begin{itemize}
    \item \textbf{MV-DNN} \cite{elkahky2015multi}: It is one of the state-of-the-art neural network models for the cross-domain recommendation. It learns the embedding of shared entities from the ratings of the source and target domains combined using deep neural networks.
\item \textbf{EMCDR} \cite{man2017cross}: It is one of the state-of-the-art models for the full-cold-start setting.
EMCDR is a two-stage neural network model for the cross-domain setting. In the first stage, it finds the embeddings for entities by leveraging matrix factorization techniques with respect to its individual domains. In the second stage, it learns a transfer function that provides target domain embeddings for the shared entities from its source domain embeddings.
\end{itemize}
\textbf{(2) Matrix factorization based single domain models (Wide):}
\begin{itemize}
\item \textbf{PMF} \cite{mnih2008probabilistic}: Probabilistic Matrix Factorization (PMF) is a standard and well-known baseline for single domain CF.
\item \textbf{GMF}\cite{he2017neural}: Generalized Matrix Factorization (GMF) is state-of-the-art rating only single domain model proposed as part of NeuMF \cite{he2017neural}. 
\end{itemize}
\textbf{(3) Matrix factorization based CDCF models (Wide):}
\begin{itemize}
\item \textbf{GMF-CD}\cite{he2017neural}: It is same as GMF model for CDR where both source and target domain ratings are used for training.
\item \textbf{CMF} \cite{singh2008relational}: Collective Matrix Factorization (CMF) is a standard baseline for CDR when entities are shared across the domains.
\end{itemize}

\subsubsection{Parameter settings and reproducibility.} We implemented our models using Tensorflow 1.12. 
We use squared error loss as an optimization loss across all models to have a fair comparison \cite{he2017neural,lian2017cccfnet}. We use $L2$ regularization for matrix factorization models (PMF and CMF) and dropout for neural network models. Hyperparameters were tuned using cross-validation. In particular, different values used for different parameters are: $\lambda$ for $L2$ regularization from \{0.0001, 0.0005, 0.005, 0.001, 0.05, 0.01, 0.5\}, dropout from \{0.1, 0.2, 0.3, 0.4, 0.5, 0.6\}, embedding dimension (i.e., number of factors $k$) from \{8, 16, 32, 48, 64, 80\} and $\alpha$ value from \{0.05, 0.1, 0.2, 0.4, 0.6, 0.8, 0.9, 0.95\}. We train the models for a maximum of 120 epochs with early stopping criterion.
Test error corresponding to the best performing model on the validation set is reported. We randomly initialize the model parameters with normal distribution with 0 mean and 0.002 standard deviation. We adopt RMSProp \cite{goodfellow2016deep} with mini-batch optimization procedure. We tested the learning rate ($\eta$) of \{0.0001, 0.0005, 0.001, 0.005, 0.01\}, batch size of \{128, 256, 512, 1024\}. From the validation set performance, we set $\eta$ to 0.002 for PMF and CMF, 0.001 for MV-DNN, 0.005 for other models; dropout to 0.5; batch size to 512. 
Further, we obtain good performance in MV-DNN when we use 3 hidden layers with the number of neurons: $4k \rightarrow 2k \rightarrow k$ (where $k$ is the embedding dimension) and the tanh activation function; we use the framework proposed in \cite{man2017cross} for MLP part of EMCDR, that is, number of neurons: $k \rightarrow 2k \rightarrow k$ with the tanh activation function. Further, in SED, for the encoder, we use 3 hidden layers with the number of neurons: $4k \rightarrow 2k \rightarrow k$ and the sigmoid activation function, and for the decoder, we use the same as the encoder with the number of neurons interchanged, and $k$ is tuned using validation set.
\subsection{Results and Discussion}
\label{subsec:res_dis}
Tables \ref{tab:sparsity} and \ref{tab:full_cold_start} detail the comparison results of our proposed models GCMF, SED and NeuCDCF, and baseline models at different sparsity and cold-start levels on four datasets. In addition, we conduct paired t-test and the improvements of NeuCDCF over each of the compared models passed with significance value $p < 0.01$. 
The findings are summarized below.
\subsubsection{Sparsity and cold-start settings (RQ1)}
\label{subsubsec:res_dis}
As we see from Table~\ref{tab:sparsity}, NeuCDCF outperforms the other models at different sparsity levels. In particular, the performance of GCMF is better when the target domain is less sparse, but, the performance of SED improves eventually when sparsity increases. This phenomenon is more obvious on the Amazon datasets. Furthermore, adding the source domain helps in recommendation performance. This is evident from the performance of GCMF as compared to its counterparts -- PMF and GMF (single domain models).

We have two different settings for the cold-start analysis: cold-start and full-cold-start, and the performance of recommendation models is given in Tables \ref{tab:sparsity} and \ref{tab:full_cold_start}. The overall performance of NeuCDCF, in particular SED, is better than the other comparison models. 
Since SED adapts and transfers the required knowledge from source domain through source domain to target domain rating construction, it helps the cold-start entities to learn better representations.
This demonstrates that a network such as SED is important in learning the non-linear representation for cold-start entities. Note that, in cold-start settings, we use GCMF with only domain-independent embeddings. That is, we set $p_u^{\delta}$ to zero in Equation (\ref{eqn:p_G_rep}). We do this because, in full-cold-start setting, we do not have any rating for the cold-start users and the term associated with $p_u^{\delta}$ in $p_u^G$ is expected to have an appropriate learned value from the user-item interactions. Since there is no item that user $u$ has rated, its initialized value remains unchanged and this results in extra noise\footnote{\url{https://cs224d.stanford.edu/lecture_notes/notes2.pdf}}$.$ For a similar reason, we set $p_u^{\delta}$ to zero in the cold-start setting as well.
\begin{figure} 
        \begin{subfigure}[b]{0.24\textwidth}
                \centering
                \includegraphics[width=.95\linewidth]{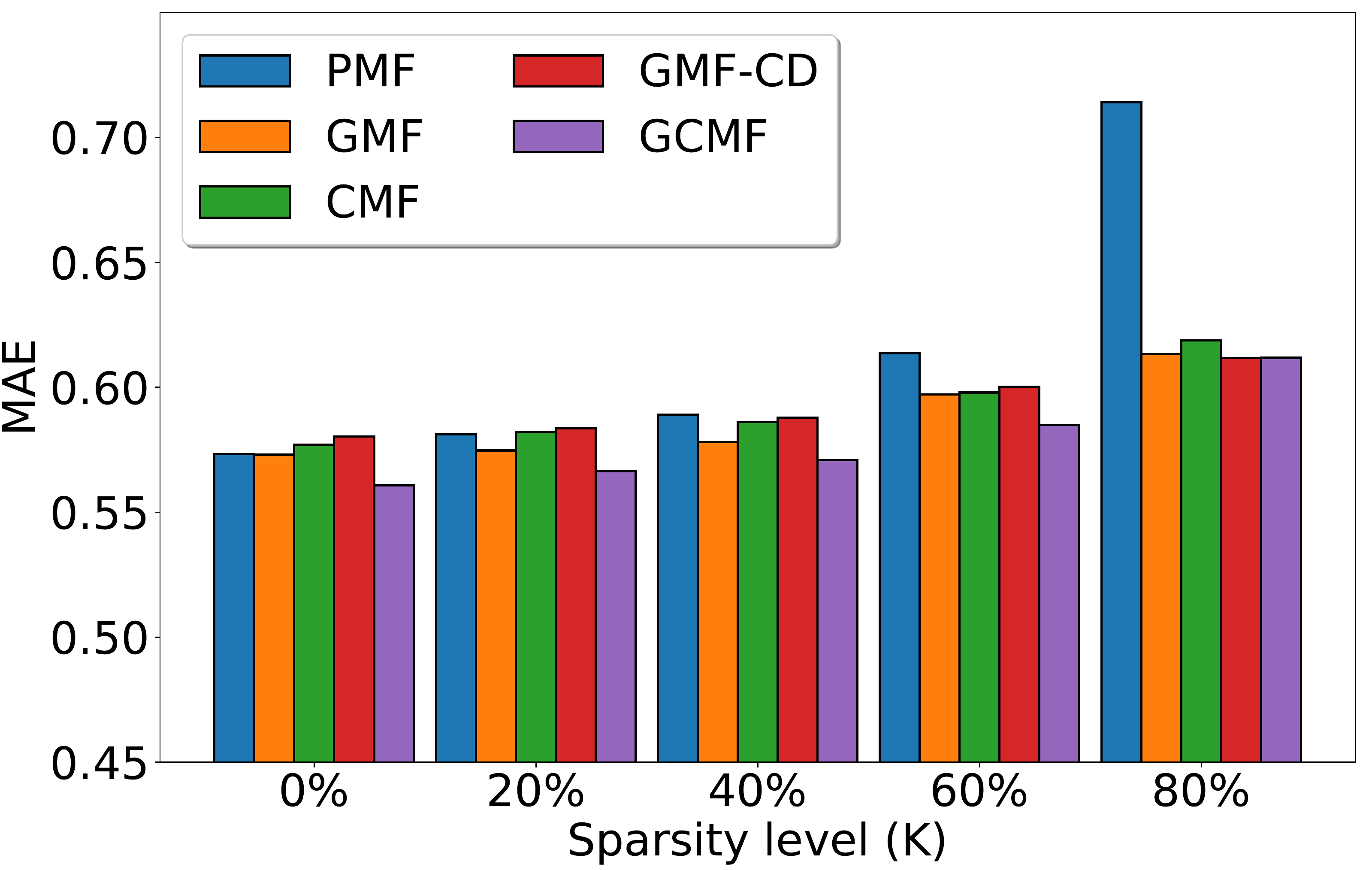}
                \caption{}
                \label{fig:ds_m_b}
        \end{subfigure}%
        \begin{subfigure}[b]{0.24\textwidth}
                \centering
                \includegraphics[width=.95\linewidth]{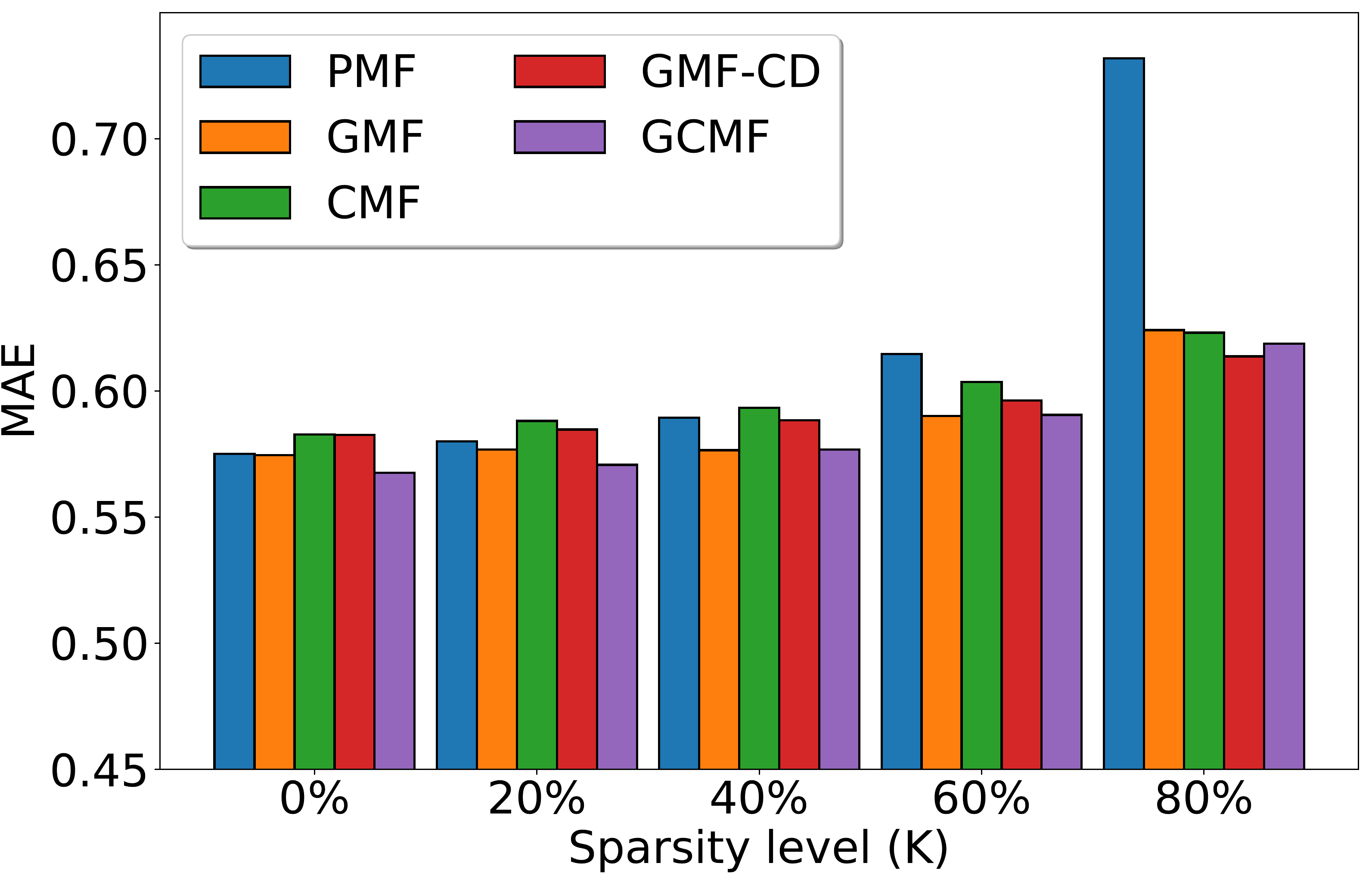}
                \caption{}
                \label{fig:ds_mu_b}
        \end{subfigure}%
        \vspace{-0.15cm} 
  \caption{Performance of GCMF model and its counterparts -- PMF, GMF, GMF-CD and CMF in MAE with respect to the various sparsity levels on (a) Douban (Movie-Book) dataset, (b) Douban (Music-Book) dataset.} \label{fig:ds}
\end{figure} 
\begin{figure} 
                \centering
                \includegraphics[width=1.0\linewidth]{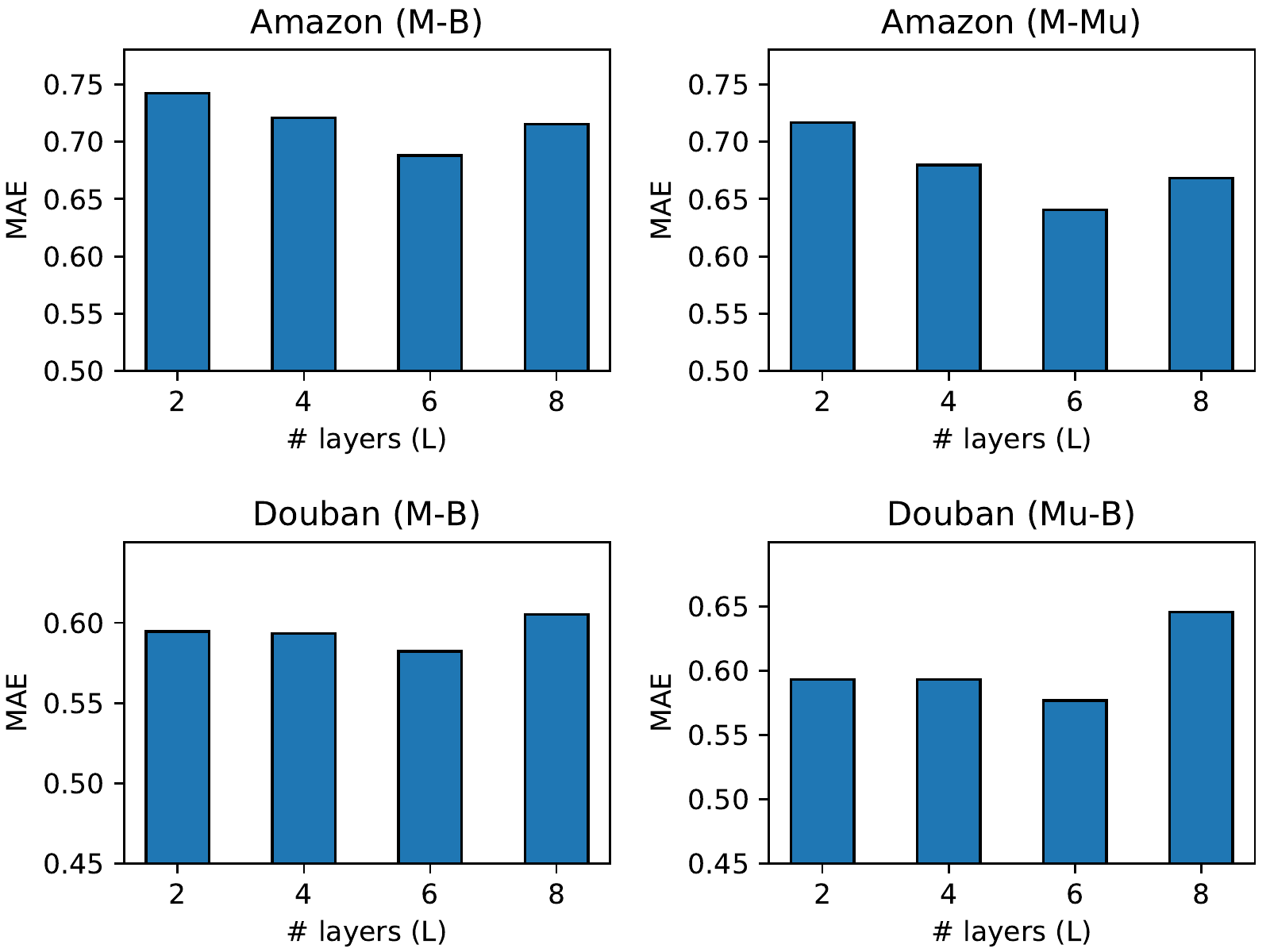}
                \label{fig:deep_layers}
        \vspace{-.35cm} 
        \caption{SED performance in MAE with respect to the number of layers (L) on all the datasets.} \label{fig:convg_2}
\end{figure}
\begin{figure}[h]
        \begin{subfigure}[b]{0.17\textwidth}
                \centering
                \includegraphics[width=.95\linewidth]{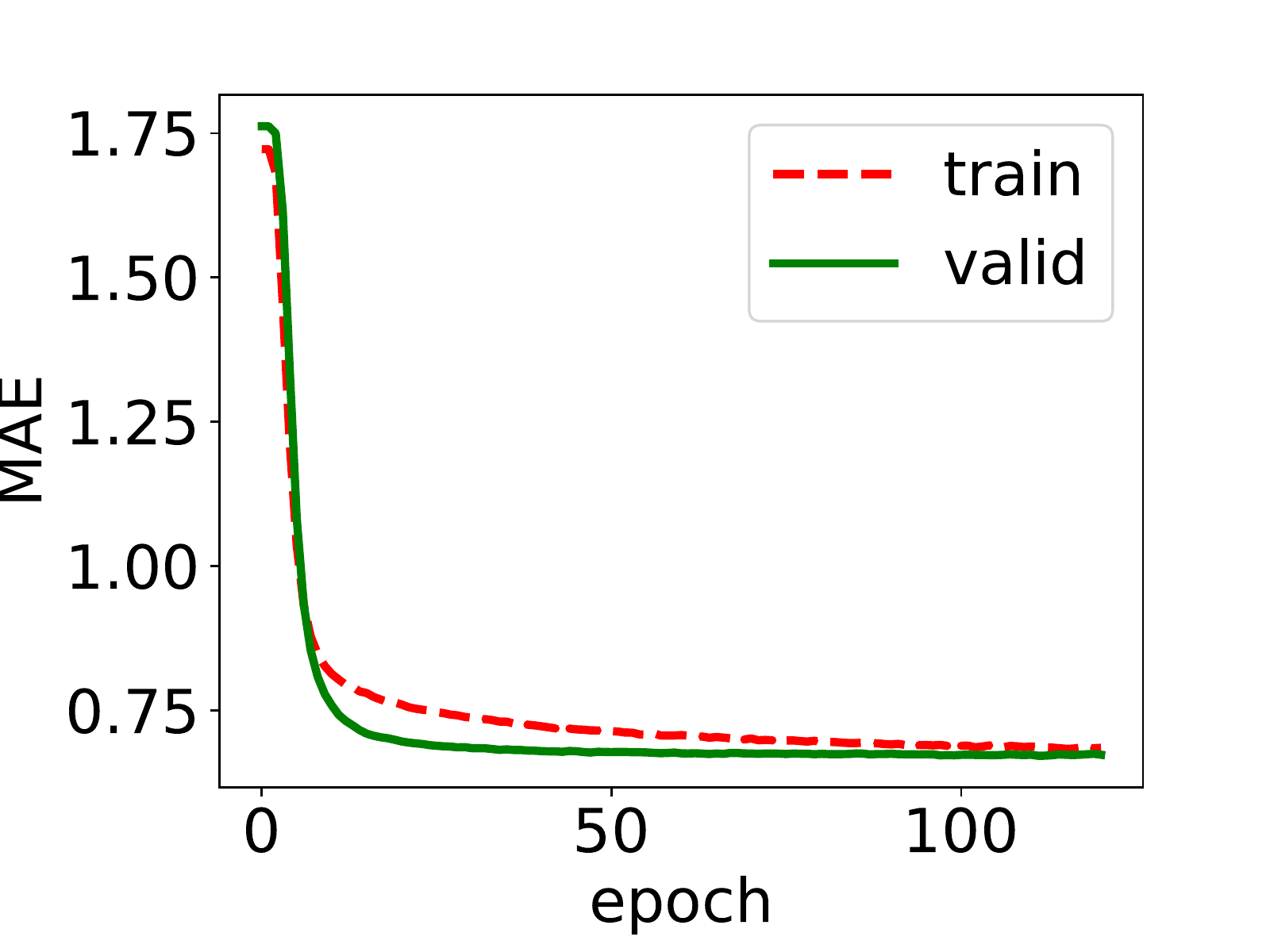}
                \caption{}
                \label{fig:convg_gcmf}
        \end{subfigure}%
        \begin{subfigure}[b]{0.17\textwidth}
                \centering
                \includegraphics[width=.95\linewidth]{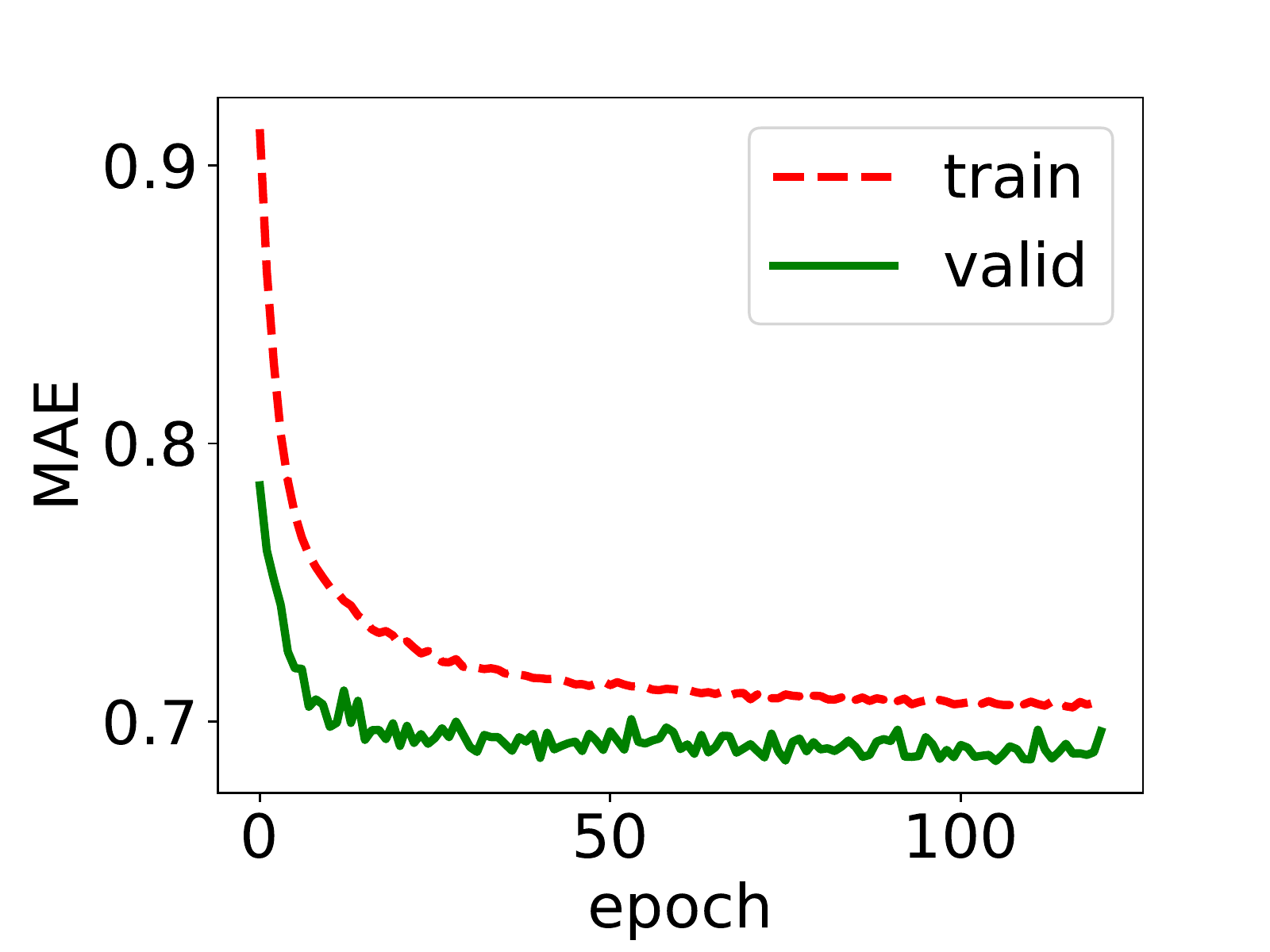}
                \caption{}
                \label{fig:convg_sed}
        \end{subfigure}%
        \begin{subfigure}[b]{0.16\textwidth}
                \centering
                \includegraphics[width=.95\linewidth]{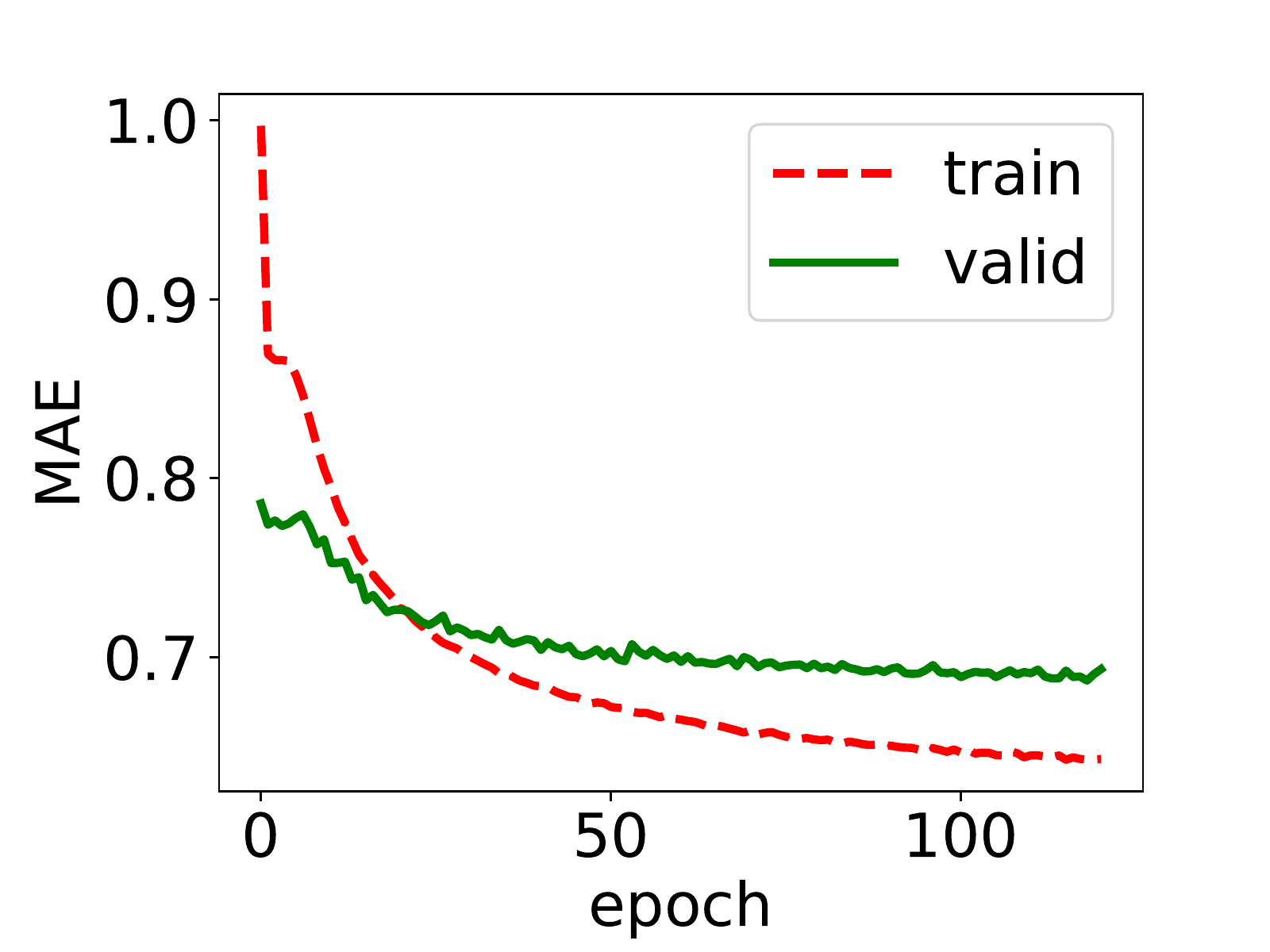}
                \caption{}
                \label{fig:convg_neucdcf_no_pretrain}
        \end{subfigure}%
        \caption{Training and validation error of (a) GCMF, (b) SED and (c) NeuCDCF with respect to the number of epochs on Amazon (Movie-Book) dataset.} \label{fig:convg_1}
\end{figure}
\subsubsection{GCMF and SED integration (RQ2)} NeuCDCF consistently performs better all the models that are either based on wide or deep framework including the proposed GCMF and SED networks. This result can be inferred from Tables \ref{tab:sparsity} and \ref{tab:full_cold_start}. 
We thus observe that the individual networks -- GCMF (wide) and SED (deep networks) provide different complementary knowledge about the relationships that exist among entities within and across domains.

\subsubsection{Domain-specific embeddings (RQ3)} Adding source domain blindly might deteriorate the performance. This is illustrated in Figures \ref{fig:ds}(a) and \ref{fig:ds}(b) using Douban datasets. When sparsity is less, surprisingly the performance of single domain models (PMF and GMF) is better than those of cross-domain models (CMF and GMF-CD). However, GCMF performs better than the counterparts -- PMF, GMF, GMF-CD and CMF in almost all the cases. These results demonstrate that GCMF provides a more principled way to understand the diversity between domains using domain-specific embeddings.

\subsubsection{Deep non-linear representations (RQ4)} Having deep representations helps in improving the performance when the sparsity increases in the target domain. This is demonstrated in Table \ref{tab:sparsity} by SED and supported by MV-DNN particularly on the Amazon datasets. Despite both being deep networks, SED performs better than MV-DNN. This is because MV-DNN learns the embeddings of shared users from source and target domain together, whereas in SED, the embeddings are learned appropriately to obtain the target domain ratings using source domain ratings.

We show the performance of SED for the different number of layers (L) on all the datasets in Figure \ref{fig:convg_2}. This shows that performance improves when we increase the number of layers ($L$) from 2 to 6. In other words, using deep representations in SED help in boosting the performance. When $L=8$ the performance decreases because of overfitting due to the high complexity of the model.

\subsubsection{Epochs needed} Figures \ref{fig:convg_1}(a), \ref{fig:convg_1}(b) and \ref{fig:convg_1}(c) show the training and validation error of GCMF, SED and NeuCDCF with respect to the number of epochs on the Amazon (Movie-Book) dataset, respectively. Behavior on the other datasets is similar, hence, we omit the details.\vspace{-0.2cm}
\section{Related Work}
\label{sec:rel}
In the literature of CDR, early works \cite{singh2008relational,li2009can,li2009transfer,gao2013cross, li2014matching} mainly adopt matrix factorization models. In particular, \cite{li2009can} constructs a cluster-level rating matrix (code-book) from user-item rating patterns and through which it establishes links to transfer the knowledge across domains. A similar approach with an extension to soft-membership was proposed in \cite{li2009transfer}. Collective matrix factorization (CMF) \cite{singh2008relational} was proposed for the case where the entities participate in more than one relation. 
However, as many studies pointed out, MF models may not handle non-linearity and complex relationships present in the system \cite{zhang2017deep,liu2015non,he2017neural}.

On the other hand, recently, there has been a surge in methods proposed to explore deep learning networks for recommender systems \cite{zhang2017deep}. Most of the models in this category focus on utilizing neural network models for extracting embeddings from side information such as reviews \cite{zheng2017joint}, descriptions \cite{kim2016convolutional}, content information \cite{wang2015collaborative}, images \cite{he2016vbpr} and knowledge graphs \cite{zhang2016collaborative}. Nevertheless, many of these models are traces to matrix factorization models, that is, in the absence of side information, these models distill to either MF \cite{koren2009matrix} or PMF \cite{mnih2008probabilistic}.

More recently, to combine the advantages of both matrix factorization models and deep networks such as multi-layer perceptron (MLP), some models have been proposed \cite{cheng2016wide,he2017neural,xue2017deep} for learning representations from only ratings. These models combine both the wide and deep networks together to provide better representations.
Autoencoder, stacked denoising autoencoder \cite{sedhain2015autorec,strub2015collaborative,wu2016collaborative,li2017collaborative}, Restricted Boltzmann machines \cite{salakhutdinov2007restricted} and recurrent neural networks \citep{ko2016collaborative,wu2017recurrent} have also been exploited for recommendation systems. 
However, the above neural network models use only the interaction between users and items from a single domain. Hence, they suffer from the aforementioned issues such as sparsity and cold-start.

Though the use of multiple related domains and neural networks for recommendations has been studied and justified in many works \cite{zhang2017deep}, very few attempts have been made to make use of neural networks in cross-domain recommendation setting \cite{elkahky2015multi,lian2017cccfnet,man2017cross,zhu2018deep}. In particular, MV-DNN \cite{elkahky2015multi} uses an MLP to learn shared representations of the entities participating in multiple domains. A factorization based multi-view neural network was proposed in CCCFNet \cite{lian2017cccfnet}, where the representations learned from multiple domains are coupled with the representations learned from content information. A two-stage approach was followed in \cite{man2017cross,zhu2018deep}, wherein the first stage, embeddings are learned for users, and in the second stage, a function is learned to map the user embedding in the target domain from the source domain.

While the models \cite{elkahky2015multi,lian2017cccfnet,man2017cross,zhu2018deep} consider learning embeddings together, they completely ignore the domain-specific representations for the shared users or items. The performance of these models \cite{elkahky2015multi,lian2017cccfnet} is heavily dependent on the relatedness of the domains. In contrast, our proposed model learns domain-specific representations that significantly improves the prediction performance. Further, \cite{lian2017cccfnet} rely on content information to bridge the source and target domains. Besides, all of these models \cite{elkahky2015multi,lian2017cccfnet,man2017cross} are either based on wide or deep networks but not both.
We are also aware of the models proposed for cross-domain settings \cite{kanagawa2018cross,xin2015cross,ma2018your,lian2017cccfnet}. However, they differ from the research scope of ours because they bridge the source and target domains using available side information. 
\section{Conclusion}
\label{sec:con}
In this paper, we proposed a novel end-to-end neural network model, NeuCDCF which is based on wide and deep framework. NeuCDCF addresses the main challenges in CDR -- diversity and learning complex non-linear relationships among entities in a more systematic way. Through extensive experiments, we showed the suitability of our model for various sparsity and cold-start settings 

The proposed framework is general and can be easily extended to a multi-domain setting as well as the setting where a subset of entities is shared. Further, it is applicable to ordinal and top-N recommendation settings with the only modification in the final loss function. NeuCDCF is proposed for rating only settings when no side information is available. If side information is available, it can easily be incorporated as a basic building block in place of matrix factorization to extract effective representations from user-item interactions. 
Despite being simple, the proposed encoder-decoder network provides significant performance improvement. 
Therefore, we believe that our work will open up the idea of utilizing more complex, at the same time, scalable encoder-decoder networks in the wide and deep framework to construct target domain ratings in the cross-domain settings.
\bibliographystyle{ACM-Reference-Format}
\bibliography{neucdcf}
\end{document}